\documentclass[prd,twocolumn]{revtex4-1}
\usepackage{amsfonts,amsmath,amssymb}
\usepackage{graphicx}
\usepackage{xcolor}

\newcommand{ \eq}[1]{Eq.~(\ref{eq:#1})}
\newcommand{\be}{\begin{equation}}
\newcommand{\ee}{\end{equation}}

\begin{document}
%------------------------------- TITLE -----------------------------%
\title{\boldmath New physics at the MUonE experiment at CERN}

\author{A.\ Masiero}
\affiliation{Dipartimento di Fisica e Astronomia `G. Galilei', Universit\`a di Padova, Italy}
\affiliation{Istituto Nazionale Fisica Nucleare, Sezione di Padova, Padova, Italy}

\author{P.\ Paradisi}
\affiliation{Dipartimento di Fisica e Astronomia `G. Galilei', Universit\`a di Padova, Italy}
\affiliation{Istituto Nazionale Fisica Nucleare, Sezione di Padova, Padova, Italy}

\author{M.~Passera} 
\affiliation{CERN, Theoretical Physics Department, Geneva, Switzerland}
\affiliation{Istituto Nazionale Fisica Nucleare, Sezione di Padova, Padova, Italy}

%------------------------------- ABSTRACT --------------------%
\begin{abstract} 
\noindent 
A confirmation of the long-standing muon $g$-2 discrepancy requires both experimental and theoretical progress. On the theory side, the hadronic corrections are under close scrutiny, as they induce the leading uncertainty of the Standard Model prediction. Recently, the MUonE experiment has been proposed at CERN to provide a new determination of the leading hadronic contribution to the muon $g$-2 via the measurement of the differential cross section of muon-electron scattering. The precision expected at this experiment raises the question whether possible new physics (NP) could affect its measurements. We address this issue studying possible NP signals in muon-electron collisions due to heavy or light mediators, depending on whether their mass is higher or lower than ${\cal O} (1{\rm GeV})$. We analyze the former in a model-independent way via an effective field theory approach, whereas for the latter we focus on scenarios with light scalar and vector bosons. Using existing experimental bounds, we show that possible NP effects in muon-electron collisions are expected to lie below MUonE's sensitivity. This result confirms and reinforces the physics case of the MUonE proposal.
\end{abstract} 

\date{\today}
\maketitle

%%%%%%%%%%%%%%%%%%%%%%%%%%%%%%%%%%%%%%%%%%%%%%%%%%%%%%%%
\section{Introduction}  \label{Intro}
%%%%%%%%%%%%%%%%%%%%%%%%%%%%%%%%%%%%%%%%%%%%%%%%%%%%%%%%

The long-standing muon $g$-2 discrepancy is one of the most intriguing hints of New Physics (NP) emerged so far in particle physics. On the experimental side, the new E989 Muon $g$-2 experiment is presently running at Fermilab and is expected to improve the current precision by a factor of four~\cite{Grange:2015fou}. In addition, a completely new low-energy approach to measuring the muon $g$-2 is being developed by the E34 collaboration at J-PARC~\cite{J-PARC}. On the theory side, considerable effort is being expended to reduce the uncertainty in the Standard Model (SM) prediction, which is dominated by the hadronic corrections.

The leading order hadronic contribution to the muon $g$-2, $a_{\mu}^{\rm HLO}$, has been traditionally computed via a dispersion integral using hadronic production cross sections in electron-positron annihilation at low energies~\cite{Keshavarzi:2019abf, Davier:2019can, Jegerlehner:2017gek}. Alternative evaluations of $a_{\mu}^{\rm HLO}$ can be obtained via lattice QCD calculations~\cite{Meyer:2018til}. 
A few years ago, a novel approach has been proposed to determine $a_{\mu}^{\rm HLO}$ measuring the leading hadronic contribution to the effective electromagnetic coupling, $\Delta \alpha_{\rm h} (q^2)$, for spacelike squared four-momentum transfers $q^2 = t <0$, via scattering data~\cite{Calame:2015fva}. The elastic scattering of high-energy muons on atomic electrons was then identified as an ideal process for this measurement, leading to the proposal of the MUonE experiment at CERN to extract $\Delta \alpha_{\rm h} (t)$ from the $\mu e$ scattering differential cross section~\cite{Abbiendi:2016xup}. For this new $a_{\mu}^{\rm HLO}$ determination to be competitive, the shape of the $\mu e$ differential cross section must be measured with a systematic uncertainty of $\mathcal{O} \left(10^{-5} \right)$, or better, close to the kinematic end point~\cite{LoI}.

In order to extract $\Delta \alpha_{\rm h} (t)$ from MUonE's precise $\mu  e$ scattering data, possible contaminations from NP effects must lie below the expected experimental resolution of $\mathcal{O} \left(10^{-5} \right)$. At the energy scale of the MUonE experiment, of ${\cal O} (1{\rm GeV})$, the leading order (LO) QED prediction for the $\mu  e$ scattering differential cross section $d\sigma_0/dt$, due to the $t$-channel exchange of a photon, dominates the SM prediction. At these energies, the weak interactions can be described by the Fermi theory and their leading correction to $d\sigma_0/dt$ is $|\delta_{Z}| \sim |t \, G_F/4\pi\alpha\sqrt{2}|\lesssim 10^{-5}$, where $G_F$ and $\alpha$ are the Fermi and fine-structure constants. As this correction is barely within MUonE's reach, the SM weak contribution to $\mu e$ scattering can be viewed as a benchmark to establish whether NP effects can be visible at the MUonE experiment. If NP lies at a scale $\Lambda \sim{\rm TeV}$, we expect that $\delta_{\rm NP} /\delta_{Z} \!\sim G_{\rm NP}/G_{F}$, where $G_{\rm NP}\sim g^{2}_{\rm NP}/\Lambda^2$ and $g_{\rm NP}$ is a typical NP coupling. As a result, $|\delta_{\rm NP}| \gtrsim 10^{-5}$ typically implies a strongly coupled NP sector. Nevertheless, NP effects of electroweak size are not implausible. In fact, the observed muon $g$-2 discrepancy 
$\Delta a_\mu=a_\mu^{\rm EXP}-a_\mu^{\rm SM}=280(74)\times 10^{-11}$~\cite{BNL,Keshavarzi:2019abf} 
can be accommodated invoking a NP effect of the same size as the SM weak contribution $\sim 5\,G_{F} M^2 / 24\sqrt{2}\pi^2$, where $M$ is the muon mass. It is therefore crucial to understand whether a NP contribution able to solve the muon $g$-2 anomaly is also polluting the extraction of $\Delta \alpha_{\rm h}(t)$ at the MUonE experiment.

In this article we consider possible signals of NP in $\mu e$ collisions due to heavy or light mediators, depending on whether their mass is higher or lower than ${\cal O} (1{\rm GeV})$, the energy scale of the MUonE experiment. In the former case, we employ an effective field theory (EFT) formalism focusing on the most general effective Lagrangian invariant under the electromagnetic gauge group.
In order to evaluate the low-energy predictions of this Lagrangian, we take into account the running effects from the NP scale $\Lambda$ down to ${\cal O} (1{\rm GeV})$ by using standard renormalisation group equation (RGE) techniques. After calculating the NP corrections to the $\mu^{\pm} e^- \!\to\! \mu^{\pm} e^-$ differential cross section, we evaluate the correlated corrections to the total cross section and forward-backward asymmetry of the process $e^+e^- \!\to \mu^+\mu^-$, in order to establish the room left to NP in $\mu e$ scattering once the experimental bounds on these observables are taken into account. Moreover, we show that four-lepton operators can generate leptonic dipoles, in particular the electron $g$-2 and electric dipole moment (EDM) which are tightly constrained experimentally. We conclude our heavy mediator analysis studying lepton flavor violating (LFV) effects in $\mu e$ collisions such as the process $\mu^+ e^- \to \mu^- e^+$, which is correlated with the muonium-antimuonium oscillation.

We finally turn our attention to light NP mediators, which cannot be analyzed in the same model-independent fashion employed for the heavy NP ones. Here we focus on popular scenarios containing either light (pseudo)scalars, referred to as axionlike particles (ALPs), or light (axial)vector bosons, such as the so-called dark photons and light $Z'$. Using existing direct and indirect bounds on masses and couplings of these light particles, we establish the maximum sizes of these light NP effects allowed in $\mu e$ collisions.

%%%%%%%%%%%%%%%%%%%%%%%%%%%%%%%%%%%%%%%%%%%%%%%%%%%%%%%%
\section{SM cross section and experimental sensitivity} \label{SM}
%%%%%%%%%%%%%%%%%%%%%%%%%%%%%%%%%%%%%%%%%%%%%%%%%%%%%%%%

The leading order SM prediction for the differential cross section of the elastic scattering $\mu^{\pm} e^- \to \mu^{\pm} e^- $ is
\begin{align}
	\frac{d \sigma_{\rm LO}^{\pm}}{dt}  =  & \,\,
	\frac{d \sigma_0}{dt}  \left( 1+ \delta_Z^{\pm} \right), 
\label{eq:sigmaLO} \\
	\frac{d \sigma_0}{dt}  =  & \,\, 
	\frac{ 4 \pi \alpha^2 f(s,t)}{t^2 \lambda \left(s,M^2,m^2 \right)} \, ,
\label{eq:sigmaLOQED} \\
	\delta_Z^{\pm} = & -\frac{G_F \, t}{4 \pi \alpha \sqrt 2} 
	\left[ a_{\theta}^2 \pm \frac{\left(s-u\right) t}{2f(s,t)} \right],
\label{eq:deltaZ}
\end{align}
where $m$ ($M$) is the electron (muon) mass, $\{s, t, u\}$ are the Mandelstam variables satisfying $s+t+u=2m^2+2M^2$, $a_{\theta}=4 s^2_{\theta} -1$, $s^2_{\theta} \approx 0.22$ is the squared sine of the weak mixing angle, $\lambda(x,y,z)=x^2+y^2+z^2-2xy-2xz-2yz$ is the K{\"a}ll{\'e}n function, and 
\be
	f(s,t)  =  t^2/2 + st + \left(M^2+m^2-s\right)^2. 
\ee	
Equation~(\ref{eq:sigmaLOQED}) is the LO QED prediction, while the term $\delta_Z^{\pm}$ is the LO correction induced by the exchange of a $Z$ boson for $|q^2| \ll M_Z^2$.

In a fixed-target experiment where the electron is initially at rest, $E_{\mu}$ is the energy of the incoming muons or antimuons, and $E$ is the electron recoil energy, the Mandelstam variables $s$ and $t=q^2$ are given by
\begin{align}
	&s \, = \,  2mE_{\mu}+M^2+m^2, \\
	&t  \, = \, -2m(E-m), \\
	& t_{\rm min} < t < 0, \quad t_{\rm min} = -\lambda(s,M^2,m^2)/s.
\end{align}	
It is also convenient to define the variable
\be
	x(t)=\left( 1- \beta \right) (t/2M^2),
\ee
with 
$\beta = (1-4M^2/t)^{1/2}$. For $E_\mu=150$~GeV, which is a typical energy available at the M2 beam line in CERN's North Area, 
$s = 0.164$ GeV$^2$, $-0.143 ~ {\rm GeV}^2 < t < 0$ and $0<x<0.932$. For these values of $s$ and $t$, the $Z$ boson correction 
	$\delta_Z^+$ is negative,
	$\delta_Z^-$ is positive,
and
	$0 < |\delta_Z^{\pm}|  < 1.5 \times10^{-5}$, 
with 
	$\delta_Z^{\pm}=0$ for $t=0$. 
As the MUonE experiment is expected to measure the shape of the differential cross section with a relative uncertainty of ${\cal O}(10^{-5})$ or better close to the end point $t = t_{\rm min}$, the maximum $Z$ boson effect is expected to be comparable with the experimental uncertainty. The tiny correction to \eq{sigmaLO} induced by the exchange of a Higgs boson of mass $ M_H$ is further suppressed by a factor of ${\cal O}(m^2 M^2 / t M^2_H)$ with respect to $\delta_Z^{\pm}$ and is therefore negligible.

Next-to-leading order (NLO) QED corrections to \eq{sigmaLOQED} were computed long time ago in~\cite{Nikishov:1961,Eriksson:1961,Eriksson:1963,VanNieuwenhuizen:1971yn,DAmbrosio:1984abj,Kukhto:1987uj,Bardin:1997nc}, with various approximations, and revisited in~\cite{Kaiser:2010zz}. The complete calculation of the full set of NLO QED and electroweak corrections, with the development of a fully differential fixed order Monte Carlo code, was completed in~\cite{Alacevich:2018vez}. The next-to-next-to-leading order (NNLO) QED corrections to $\mu e $ scattering are under investigation~\cite{Mastrolia:2017pfy,DiVita:2018nnh,Mastrolia:2018sso,DiVita:2019lpl,Engel:2018fsb, Engel:2019nfw}. The resummation of classes of higher order QED corrections enhanced by large logarithms will be mandatory to match MUonE's extremely high accuracy~\cite{Abbiendi:2016xup}. The leading hadronic corrections to \eq{sigmaLOQED} are given by 
\be
	\frac{d \sigma_{\rm NLO, h}^{\pm}}{dt}  \,=\,  2 \Delta \alpha_{\rm h} (t) \, \frac{d \sigma_0}{dt}.
\label{eq:sigmaHAD}
\ee
For $E_{\mu} = 150$~GeV, $2\Delta \alpha_{\rm h} (t)$ reaches the maximum value of $2.1 \times 10^{-3}$ at $t=t_{\rm min}=-0.143~{\rm GeV}^2$. The NNLO hadronic corrections to $\mu e$ scattering were recently computed in~\cite{Fael:2018dmz,Fael:2019nsf}.

The collision of positive muons and electrons can also lead to the production of neutrino-antineutrino pairs via the process $\mu^{+} e^- \to \nu_e \bar{\nu}_{\mu}$. For squared four-momentum transfers much smaller, in absolute value, than the squared $W$-boson mass, this SM cross section is, neglecting terms of $\mathcal{O} (m^2/s)$,
$ G_F^2 \left( M^2 + 2s \right) /12 \pi$.
For $E_\mu=150$~GeV, it leads to a tiny $4.8 \times 10^{-10} \mu$b, while the LO QED elastic cross section for the same value of $E_\mu$ and $E>1$~GeV is $\sigma_0 = 245 \mu$b. This process is therefore negligible at the MUonE experiment.

The MUonE experiment is expected to determine $\Delta \alpha_{\rm h} (t)$ in a kinematic region relevant to calculate the leading hadronic contribution to the muon $g$-2. In particular, this quantity will be extracted from the shape of the differential $\mu e$ scattering cross section by a template fit method~\cite{LoI}. The basic idea is that $\Delta \alpha_{\rm h} (t)$ can be obtained measuring, bin by bin, the ratio $(N_i/N_{\rm {n}})^{\pm}$ (as earlier, the superscript $\pm$ refers to $\mu^{\pm}$ beams), where $N_i$ is the number of scattering events in a specific $t$-bin, labeled by the index $i$, and $N_{\rm {n}}$ is the number of events in the normalization $t$-bin corresponding to $x(t) \sim 0.3$ (for this value of $x$, $\Delta \alpha_{\rm h} (t)$ is comparable to the experimental sensitivity expected at MUonE and its error is negligible). Therefore, this measurement will not rely on the absolute knowledge of the luminosity. To extract the leading hadronic corrections to the $\mu e$ scattering cross section in the $t$-bin $i$, let us split the theoretical prediction into 
\be
	\sigma^{\pm}_{\rm {TH},i} = \sigma_{0,i} \left[ 1 + 2 \Delta \alpha_{{\rm h},i} + 
		\delta^{\pm}_{i} + \delta^{\pm}_{{\rm NP},i} \right],
\ee
where $\sigma_{0, i} = \int_i (d \sigma_0 / dt ) dt$ is the LO QED prediction obtained integrating~\eq{sigmaLOQED} in the $t$-bin $i$, $2 \Delta \alpha_{{\rm h},i}$ is the leading hadronic correction obtained from~\eq{sigmaHAD}, $\delta_i^{\pm}$ is the remainder of the SM corrections, and $\delta^{\pm}_{{\rm NP},i}$ is a possible NP contribution. The experimentally measured ratio $(N_i/N_{\rm {n}})^{\pm}$ can then be equated with the ratio of the theoretical predictions,
\begin{align}
	\left( \frac{N_i}{N_{\rm {n}}} \right)^{\pm} & \!\! = 
%	\,\,
	 \left( \frac{\sigma_{\rm {TH},i}}{\sigma_{\rm {TH,n}}} \right)^{\pm} \!\!
		\simeq 
%		\,\,
	  \frac{\sigma_{\rm {0},i}}{\sigma_{\rm {0,n}}} \, \Big[ 
		1+2 \left( \Delta \alpha_{{\rm h},i} - \Delta \alpha_{{\rm h,n}}\right)  
		\nonumber\\[1mm]		 
	& +
		 \left(\delta_i -\delta_{\rm n} \right)^{\pm}+ 
		 \left(\delta_{{\rm NP}, i} - \delta_{\rm NP,n} \right )^{\pm} \Big] .
\label{eq:norm}
\end{align}
As $\Delta \alpha_{{\rm h,n}}$ is known with negligible error, if $\left(\delta_i -\delta_{\rm n} \right)^{\pm}$ is computed with sufficient precision, one can extract $2 \Delta \alpha_{{\rm h},i} + \left(\delta_{{\rm NP},i} -\delta_{\rm NP,n} \right )^{\pm}$, bin by bin, from $(N_i/N_{\rm {n}})^{\pm}$. Equation~(\ref{eq:norm}) shows that the impact of the SM corrections on this extraction can only be established after subtracting their value in the normalization region. From \eq{norm} we can also conclude that the MUonE experiment will not be sensitive to a NP signal constant in $t$ relative to the LO QED one, i.e.\  such that $\delta_{{\rm NP},i}  = \delta_{\rm NP,n}$.

%%%%%%%%%%%%%%%%%%%%%%%%%%%%%%%%%%%%%%%%%%%%%%%%%%%%%%%%
\section{Heavy New physics mediators} \label{Heavy}
%%%%%%%%%%%%%%%%%%%%%%%%%%%%%%%%%%%%%%%%%%%%%%%%%%%%%%%%

In this Section we consider possible heavy NP effects in low-energy collisions of positive and negative muons with electrons. The masses $\Lambda$ of the mediators are assumed to be much larger than ${\cal O} (1{\rm GeV})$, so that an EFT approach is appropriate to encode the leading NP contributions.

%%%%%%%%%%%%%%%%%%%%%%%%%%%%%%%%%%%%%%%%%%%%%%%%%%%%%%%%
\subsection{Effective Lagrangian}
\label{section:effectivelagrangian}

The most general effective Lagrangian for charged leptons invariant under the electromagnetic gauge group can be written, 
up to dimension-6 effective operators, as~\cite{Jenkins:2017jig}
\begin{align}
	\mathcal{L}_{\rm{LEFT}} \, = \,
           \big[	
        & d^{prst}_{1} \left( \bar{e}_{pL} e_{rR} \right) \!\left( \bar{e}_{sL} e_{tR} \right) + h.c. \, +
	\nonumber\\
	& d^{prst}_{2} \left( \bar{e}_{pL} \gamma^\mu e_{rL} \right) \!\left( \bar{e}_{sL} \gamma^\mu e_{tL} \right) + 
	\nonumber\\
	& d^{prst}_{3} \left( \bar{e}_{pL} \gamma^\mu e_{rL} \right) \!\left( \bar{e}_{sR} \gamma^\mu e_{tR} \right) + 
	\nonumber\\
	& d^{prst}_{4} \left( \bar{e}_{pR} \gamma^\mu e_{rR} \right) \!\left( \bar{e}_{sR} \gamma^\mu e_{tR} \right) 
	 \big] / \Lambda^2 \, +
	 \nonumber\\
	 & d^{pr}_{0} \left( \bar{e}_{pL} \,\sigma^{\mu\nu} e_{rR} \right) F_{\mu\nu}/ \Lambda + h.c. \, ,
 \label{eq:Leff_LEFT} 
\end{align}
where $p,r,s,t$ are flavor indices. The dipole operators can contribute to $\mu e$ scattering by means of a double operator insertion. However, taking into account the tight experimental constraints on leptonic dipole moments, we find that they can be safely neglected in our study. Similarly, we neglected four-fermion semileptonic operators as they can contribute to $\mu e$ scattering only at the one-loop level through the generation of four-lepton operators already present in~\eq{Leff_LEFT}. In principle, also the operator $\left( \bar{\nu}_{\mu L} \gamma^\mu \mu_{L} \right) \!\left( \bar{e}_{L} \gamma^\mu \nu_{eL} \right)$ could contribute indirectly to $\mu e$ scattering, as it affects the extraction of $G_F$ from the muon decay rate. However, such a NP shift of $G_F$ is tightly constrained experimentally (for instance from $\tau$ decay data~\cite{Tanabashi:2018oca}) and it is therefore irrelevant for $\mu e$ scattering. Moreover, we remark that the Lagrangian in~\eq{Leff_LEFT} captures both the tree-level effects induced by the exchanges of heavy mediators, as well as their loop corrections (vertex corrections, vacuum polarization insertions and box effects) to the QED amplitude.

Further selecting the relevant flavor structures for $\mu e$ scattering and using Fierz identities, we obtain a lepton flavor conserving (LFC) Lagrangian 
\begin{align}
	\mathcal{L}_{\rm{LFC}}  \, = \, \big[
		    & \left( a_1 \!+\! i a_2 \right) \left( \bar{\mu}_L \mu_R \right) \!\left( \bar{e}_L  e_R \right) + h.c. \, +
						\nonumber \\
		  & \left( a_3 \!+\! i a_4 \right) \left( \bar{\mu}_L e_R \right) \!\left( \bar{e}_L  \mu_R \right) + h.c. \, +
						\nonumber \\
		  & \left( a_5 \!+\! i a_6 \right) \left( \bar{\mu}_L \mu_R \right) \!\left( \bar{e}_R  e_L \right) + h.c. \, + 	  
		 \nonumber \\
		  & ~a_7 \left( \bar{\mu}_L \gamma^{\mu} \mu_L \right) \left( \bar{e}_L \gamma_{\mu} e_L \right) +
		 				\nonumber \\
		  & ~a_8 \left( \bar{\mu}_R \gamma^{\mu} \mu_R \right) \left( \bar{e}_R \gamma_{\mu} e_R \right) +  
		   				\nonumber \\
		  & ~a_9 \left( \bar{\mu}_L \gamma^{\mu} \mu_L \right) \left( \bar{e}_R \gamma_{\mu} e_R \right) +
		 				\nonumber \\
		  & ~a_{10} \left( \bar{\mu}_R \gamma^{\mu} \mu_R \right) \left( \bar{e}_L \gamma_{\mu} e_L \right)
				   \big] / \Lambda^2, 		  
\label{eq:LHEAVY_LFC}
\end{align}
which contributes to the LFC process $\mu^{\pm} e^- \to \mu^{\pm} e^- $, and a purely lepton flavor violating (LFV) Lagrangian 
\begin{align}
	\mathcal{L}_{\rm{LFV}}   \, = \, &  \big[ \,
 	b_1 \left( \bar{\mu}_L e_R \right) \!\left( \bar{\mu}_L   e_R \right) + 
	b_2 \left( \bar{\mu}_R e_L \right) \!\left( \bar{\mu}_R  e_L \right)  +
						  \nonumber \\
	& ~~b_3 \left( \bar{\mu}_L  e_R \right) \left( \bar{\mu}_R  e_L \right) 
 		    +b_4 \left( \bar{\mu}_L \gamma^{\mu} e_L \right) \left( \bar{\mu}_L \gamma_{\mu} e_L \right) +
						  \nonumber \\						  
	& ~~b_5 \left( \bar{\mu}_R \gamma^{\mu} e_R \right) \left( \bar{\mu}_R \gamma_{\mu} e_R \right)  + {\rm h.c.}
         \big] / \Lambda^2\,,
\label{eq:LHEAVY_LFV}
\end{align}
which violates the electron and muon family numbers by two units (while preserving the total lepton number), thus generating the process $\mu^{+} e^{-} \to \mu^{-} e^{+} $. The dimensionless coefficients $a_{k}$ ($k=1,\dots,10$) are real, while $b_{l}$ ($l=1,\dots,5$) are complex.

Allowing for a more general flavor structure, we could generate the additional LFV processes $\mu^{\pm} e^- \to e^{\pm}e^-$~\cite{Koike:2010xr,Uesaka:2016-17} and $\mu^{\pm} e^- \to \mu^{\pm} \mu^-$~\cite{Ibarra}, which are however strongly constrained by the $\mu \to 3e$ and $\mu \to e \gamma$ experimental bounds~\cite{Crivellin:2013hpa,Pruna:2014asa,Crivellin:2017rmk,Calibbi:2017uvl,Lindner:2016bgg}. Therefore, since our aim is to maximise NP effects in $\mu e$ scattering, hereafter we will focus on the effective Lagrangians of Eqs.~(\ref{eq:LHEAVY_LFC}--\ref{eq:LHEAVY_LFV}).

The low-energy contributions induced by the above Lagrangians, which are defined at the scale $\Lambda\gg 1{\rm GeV}$, can be evaluated only after taking into account the running effects from the scale $\Lambda$ down to ${\cal O} (1{\rm GeV})$. By using standard RGE techniques, this procedure amounts to replacing the above Lagrangians $\mathcal{L}$ with $\mathcal{L} + \delta \mathcal{L}$, where $\delta \mathcal{L}$ stems from QED loop-induced effects. We evaluated the full one-loop expression of $\delta \mathcal{L}$ and found that the most relevant phenomenological effects stem from the dipole operators
\begin{align}
	\delta \mathcal{L}_e = \frac{e}{8\pi^2} \frac{M}{\Lambda^2} \log\left(\frac{\Lambda}{\mu}\right)
  	\left[ \, \bar{e}  \left( a_3 \!+\! i a_4 \gamma_5 \right) \sigma^{\mu\nu}\, e \, \right] F_{\mu\nu},
\label{eq:L_eff_RGE}
\end{align}
where $\mu$ is the renormalization scale and the corresponding $\delta \mathcal{L}_\mu$ for the muon is obtained from~\eq{L_eff_RGE} simply replacing $e \to \mu$ and $M \to m$~\cite{Feruglio:2018fxo}. As we will see, $\delta \mathcal{L}= \delta \mathcal{L}_e + \delta \mathcal{L}_\mu$ generates contributions to the $g$-2 and electric dipole moments of the muon and electron, so that the coefficients $a_{3,4}$ will be tightly constrained.

We remark that if NP lies above the electroweak scale, our effective Lagrangian must be invariant under the full SM gauge group~\cite{Buchmuller:1985jz, Grzadkowski:2010es} and not only under the electromagnetic $U(1)$, as assumed so far. Therefore, in this case, the conclusions that will be drawn below can be regarded as conservative. Some specific examples of heavy NP were discussed in~\cite{Schubert:2019nwm} reaching broadly the same conclusions of our general analysis presented below.

Before studying the phenomenological implications of the above Lagrangians, let us discuss the theoretical bounds arising from perturbativity and unitarity. In particular, perturbativity requires that $|a_k|, |b_l| \lesssim 16\pi^2$, where the upper bound is saturated for a maximally strong regime. Instead, the unitarity bounds read 
\be
	\frac{|a_k|,|b_l|}{\Lambda^2} <  \frac{16 \pi \, \eta_{k,l}}{s},
\label{eq:unitarity}
\ee
where $\eta_{k,l}$ are positive $\mathcal{O}(1)$ coefficients. In practice, given an experiment running at an energy $\sqrt{s}$, the Wilson coefficients of our four-fermion effective Lagrangian, i.e.\ $a_k/\Lambda^2$ and $b_l/\Lambda^2$, cannot be arbitrarily large and must satisfy the constraint of~\eq{unitarity}.

%%%%%%%%%%%%%%%%%%%%%%%%%%%%%%%%%%%%%%%%%%%%%%%%%%%%%%%%
\subsection{Heavy NP in \boldmath $\mu^{\pm} e^- \to \mu^{\pm} e^- $ scattering}
\label{section:mue_scattering}
The leading corrections induced by $\mathcal{L}_{\rm{LFC}}$ to the LO QED differential cross section $d\sigma_0/dt$ are given by
\begin{align}
	\frac{d \sigma^{\pm}_{\rm{LFC}}}{dt}  \, =  \, \frac{d \sigma_0}{dt} \, \delta^{\pm}_{\rm{LFC}},
\label{eq:deltaHEAVY}
\end{align}
where, defining $z^+=s$, $z^-=u$,
\begin{align}
	\delta^{\pm}_{\rm{LFC}} & =  \frac{t}{8 \pi \alpha \Lambda^2} \frac{1}{f(s,t)}
	\bigg[ \, 2 m M \, a_S \, \left(z^{\mp}-z^{\pm} \right)
  		\nonumber \\
	& + 2 m M \, a_T \left(z^{\pm} \!-\! t \!-\! M^2 \!-\! m^2 \right) 
		\nonumber \\
	& + \,a_V \, f(s,t) +  \frac{a_A}{2}  \, t \, \left(z^{\pm}-z^{\mp} \right) 
	\bigg]
\label{eq:deltaLFC}
\end{align}
and the coefficients $a_{S,T,V,A}$ are defined as
\begin{align}
	a_S \, &=  \, a_1 + a_5, 
		\nonumber \\
	a_T   &= \, a_3,
		\nonumber \\
	a_V \, &=	\, a_7 + a_8 + a_9 + a_{10},
		\nonumber \\
	a_A \, &=	\, a_7 + a_8 - a_9 - a_{10}.
\label{eq:theas}
\end{align}
Terms of order $(t/\Lambda^2)^2$ or suppressed by $a_{\theta}^2 \approx 10^{-2}$ were systematically neglected in~\eq{deltaLFC}.

The leading contribution induced by the exchange of a heavy scalar (pseudoscalar) mediator is obtained from Eqs.~(\ref{eq:deltaLFC}--\ref{eq:theas}) setting $a_1=a_5$ ($a_1=-a_5$) and $a_k=0$ $\forall k\neq 1,5$. Therefore, for a heavy scalar, $\delta^{\pm}_{\rm{LFC}}$ depends only on the parameter $a_S$ whereas, for a heavy pseudoscalar, $\delta^{\pm}_{\rm{LFC}}=0$. For example, for the SM Higgs boson with mass $M_H$, it is $\Lambda^2 = 1/(\sqrt{2}G_F)$ and $a_S = 2mM/M_H^2$. The leading effect of a heavy vector (axial) boson is given by Eqs.~(\ref{eq:deltaLFC}--\ref{eq:theas}) with $a_7=a_8=a_9=a_{10}$ ($a_7=a_8=-a_9=-a_{10}$) and all other $a_k=0$. Therefore, for a heavy (axial) vector, $\delta^{\pm}_{\rm{LFC}}$ depends only on the parameter  ($a_A$) $a_V$. The correction $\delta_Z^{\pm}$, \eq{deltaZ}, induced by the exchange of a $Z$-boson for $|t| \ll M_Z^2$, is obtained from Eqs.~(\ref{eq:deltaLFC}--\ref{eq:theas}) with $\Lambda^2 = 1/(\sqrt{2}G_F)$, $a_V=-(4s^2_{\theta}-1)^2$ and $a_A=-1$. The coefficient $a_T$ is introduced by spin-1 tensor interactions. Spin-2 interactions are not described by the Lagrangian $\mathcal{L}_{\rm{LFC}}$ as they require dimension-8 effective operators.

A necessary condition for NP to affect the measurements of the MUonE experiment is that they are larger than the expected experimental resolution of $\mathcal{O} \left(10^{-5} \right)$. Barring large accidental cancellations among the $a_{S,T,V,A}$ contributions to $\delta^{\pm}_{\rm{LFC}}$, at MUonE's energies this implies
\begin{align}
	&\left| a_{V,A} \right|  \gtrsim  10 \left( \! \frac{\Lambda}{1 {\rm TeV}} \! \right)^{\! 2},
	\label{eq:bound_deltaLFC}\\
	&\left| a_{S,T} \right|  \gtrsim  10^{4} \left( \! \frac{\Lambda}{1 {\rm TeV}} \! \right)^{\! 2}.
	\label{eq:bound_deltaLFC_aS_aT}
\end{align}
Pure (pseudo)scalar and tensor quadratic effects are of order $(a_{X}t/4\pi\alpha)^2/\Lambda^4$ ($X=S,P,T$) and are not suppressed by the electron and muon masses. We find that these coefficients $a_{X}$ are subject to bounds comparable to those in~\eq{bound_deltaLFC_aS_aT}.
The theoretical bounds from perturbativity and unitarity (see \eq{unitarity}), as well as the constraints imposed by the leptonic dipole moments, which will be discussed shortly, forbid any significant effect in $\mu e$ scattering arising from $a_{S,T}$. From now on we will therefore safely set $a_{S}=a_{T}=0$. Under these assumptions we note that the couplings $a_V$ and $a_A$ can be probed separately by defining, at the end point $t = t_{\rm min}$, the two observables
\begin{align}
	\!\!\!\!	   \left( \delta^{+}_{\rm{LFC}} + \delta^{-}_{\rm{LFC}} \right)_{t=t_{\rm min}} & = 
	\frac{a_V}{4 \pi \alpha \Lambda^2} \,  t_{\rm min},
\label{eq:a_V} \\ 
	\!\!\!\!\!  \left( \delta^{+}_{\rm{LFC}} - \delta^{-}_{\rm{LFC}} \right)_{t=t_{\rm min}} &= 
	\frac{a_A}{4 \pi \alpha \Lambda^2} \left( t_{\rm min} - 2M^2 + 2s \right),
\label{eq:a_A}
\end{align}
where we safely set $m=\!0$ in~\eq{a_A}. As we will see in the next Section, $a_V$ and $a_A$ are constrained 
by the cross section and forward-backward asymmetry of the process $e^+e^- \!\to \mu^+\mu^-$.

%%%%%%%%%%%%%%%%%%%%%%%%%%%%%%%%%%%%%%%%%%%%%%%%%%%%%%%%
\subsection{Heavy NP in \boldmath $e^+e^- \to \mu^+\mu^-$}

The four-lepton effective Lagrangian $\mathcal{L}_{\rm LFC}$ also contributes to the process $e^+ e^- \to \mu^+ \mu^-$. If we consider center-of-mass energies $\sqrt s$ much larger than $M$, so that both electron and muon masses can be neglected, 
the total cross section for this process is 
\begin{align}
	\sigma \left( e^+ e^- \to \mu^+ \mu^- \right) & \, =\,  \frac{4\pi \alpha^2}{3s} 
	+ \frac{\alpha G_F}{3 \sqrt 2}  \frac{a_{\theta}^2 M_Z^2}{s- M_Z^2} + \nonumber \\
	&  \!\!\!\!\!\!\!\!\!\!\!\!\!\!\!\!\!\!\!\!\!\!\!\!\!\!\!\!\!\!\!\!\!\!\!\!\! 
	+ \frac{G_{F}^2}{96 \pi} \! \left( a_{\theta}^2 +1 \right)^2 \!\! \frac{s M_Z^4}{\left( s-M_Z^2 \right)^2} \, + \nonumber \\
	&  \!\!\!\!\!\!\!\!\!\!\!\!\!\!\!\!\!\!\!\!\!\!\!\!\!\!\!\!\!\!\!\!\!\!\!\!\! 
	+ \frac{1}{\Lambda^2} \left( a_V \, \frac{\alpha}{6}  \, + \,  a_A \, \frac{G_F}{48 \pi \sqrt 2} \frac{s M_Z^2}{s- M_Z^2} \right),
\label{eq:sigmaeemumu}
\end{align}
while the forward-backward asymmetry reads
\begin{align}
	A_{\rm FB} & = A_{\rm FB}^{\rm SM} \left[ 1+ 
	\frac{r(s)}{\Lambda^2} \! \left( \frac{a_A \left( s \!-\! M_Z^2 \right)}{\sqrt{2} G_F M_Z^2} - \frac{a_V s}{16 \pi \alpha}\right) \right] \!,
\label{eq:FBAeemumu} \\
	A_{\rm FB}^{\rm SM} & =  \frac{3 s G_F M_Z^2 \left[ 4 \pi \alpha \sqrt{2} \left( s \!-\! M_Z^2 \right) 
	+ a_{\theta}^2 s G_F M_Z^2 \right] }{d(s)},
\end{align}
where the functions $r(s)$ and $d(s)$ are given by
\begin{align}
	d(s) & = 128 \pi^2 \alpha^2 \left( s- M_Z^2 \right)^2 + \left( a_{\theta}^2 +1 \right)^2 s^2 G_{F}^2 M_Z^4 +
		\nonumber \\
		& +16 \pi \alpha \sqrt{2} a_{\theta}^2 G_F M_Z^2 s \left(s - M_Z^2 \right),
		\\
	r(s) & = \frac{128 \pi^2 \alpha^2 \left( s- M_Z^2\right)^2 - s^2 G_{F}^2 M_Z^4}
	                    {128 \pi^2 \alpha^2 \left( s- M_Z^2\right)^2 + s^2 G_{F}^2 M_Z^4}.
\end{align}
Notice that in the NP contributions of Eqs.~(\ref{eq:sigmaeemumu}--\ref{eq:FBAeemumu}) we neglected terms of order $(s/\Lambda^2)^2$ 
as well as terms suppressed by $a_{\theta} = 4 s^2_\theta -1 \approx -0.1$.

The most stringent bounds on NP effects in $\sigma(e^+e^- \to \mu^+\mu^-)$ and $A_{\rm{\rm FB}}$ are set by the LEP-II data~\cite{Schael:2013ita}:
\begin{align}
	\frac{\sigma(e^+e^- \!\to\! \mu^+\mu^-)_{\rm {\rm EXP}}}{\sigma(e^+e^- \!\to\! \mu^+\mu^-)_{\rm{\rm SM}}} & = 0.9936 \pm 0.0141\,,
\label{eq:exp_LEPs}
		\\
	\frac{A_{\rm{\rm FB}}^{\rm {\rm EXP}}(e^+e^- \!\to\! \mu^+\mu^-)}{A_{\rm{\rm FB}}^{\rm {\rm SM}}(e^+e^- \!\to\! \mu^+\mu^-)} & 
	= 0.9925 \pm 0.0212\,,
\label{eq:exp_LEPA}
\end{align}
where the ratios refer to the mean values in the energy range $130 \leq \sqrt{s} \leq 207~$GeV. Imposing the above experimental bounds at the $2\sigma$ level, we find that 
\begin{align}
	| a_{V,A} | & \lesssim \left( \! \frac{\Lambda}{1 {\rm TeV}} \! \right)^{\! 2}
\end{align}
for $\Lambda$ much larger than the LEP-II energies, which is not compatible with the requirement of visible NP effects in $\mu e$ scattering, see~\eq{bound_deltaLFC}.

We now turn to possible NP at or below the electroweak scale. Since the EFT approach is valid as long as $\sqrt{s} \ll \Lambda$, LEP-II data can no longer be used for this analysis. However, we can still rely on low-energy data from PEP~\cite{Derrick:1985gs}, PETRA~\cite{Hegner:1989rd}, and TRISTAN~\cite{Miura:1997mq}, which ran at the center of mass energies $\sqrt{s}=(29,35,58)~$GeV, respectively. The measured values of the $e^+ e^- \to \mu^+ \mu^-$ cross sections are
\begin{align}
\!\frac{\sigma (e^+ \! e^- \!\to\! \mu^+ \! \mu^-)_{\rm EXP}}{\sigma (e^+ \! e^- \!\to\! \mu^+ \! \mu^-)_{\rm SM}}  =\! 
\Bigg\{
	\begin{array}{ll}
        \!0.994 \pm 0.022,~  \sqrt{s}=29~{\rm GeV} \\
	    \!0.984 \pm 0.027,~ \sqrt{s}=35~{\rm GeV}  \\
     	\!0.987 \pm 0.019,~ \sqrt{s}=58~{\rm GeV}
	\end{array}\,,
\nonumber
\end{align}
while for the forward-backward asymmetry they found
\begin{align}
	\!\frac{A_{\rm FB}^{\rm EXP}  (e^+ \! e^- \!\to\! \mu^+ \! \mu^-)  }{A_{\rm FB}^{\rm SM} (e^+ \! e^- \!\to\! \mu^+ \! \mu^-) }  =\! 
	\Bigg\{
	\begin{array}{ll}
  	\!0.995 \pm 0.164,~ \sqrt{s} =29~{\rm GeV} \\
  	\!1.076 \pm 0.170,~ \sqrt{s}=35~{\rm GeV}  \\
  	\!0.977 \pm 0.065,~ \sqrt{s}=58~{\rm GeV}
	\end{array}\,. 
	\nonumber
\end{align}
If $s \ll M_Z^2$, the expressions in Eqs.~(\ref{eq:sigmaeemumu}--\ref{eq:FBAeemumu}) can be approximated by
\begin{align}
	& \sigma \left( e^+ \! e^- \!\to\! \mu^+ \! \mu^- \right) \approx  \frac{4\pi \alpha^2}{3s} 
	+ \frac{\alpha}{6} \frac{a_V}{\Lambda^2},
	\\
	& A_{\rm FB}  \approx  A_{\rm FB}^{\rm SM} \left( 1- \frac{a_A }{\sqrt{2} G_F\Lambda^2} \right),
\label{eq:eemumu_low_energy}
\end{align}
showing that, in this approximation, $a_A$ and $a_V$ are separately probed by $A_{\rm FB}$ and $\sigma \left( e^+ e^- \to \mu^+ \mu^- \right)$. We find that PEP data imply the bound $\delta^{\pm}_{\rm LFC} \lesssim 10^{-5}$, which is trustable only for $\Lambda \gtrsim 100~$GeV since the EFT approach breaks down for lower values of $\Lambda$. 
 
On the other hand, new particles with masses below~$\sim 100$ GeV are disfavored by direct searches at LEP. Moreover, at tree level, the effective couplings $a_V$ and $a_A$ can only be induced by the exchange of a vector boson $U$ with the vector and axial-vector couplings to leptons $g_V^\ell \, \bar\ell \gamma^\mu \ell \, U_\mu$ and $g_A^\ell \, \bar\ell \gamma^\mu \gamma_5 \ell \, U_\mu$ ($\ell =e,\mu$). In this case, Eqs.~(\ref{eq:sigmaeemumu}--\ref{eq:FBAeemumu}) can still be used replacing $a_X/\Lambda^2 \to 4 g^e_Xg^\mu_X/(s- M^2_U)$, where $X=V,A$. Imposing the LEP-II bounds of Eqs.~(\ref{eq:exp_LEPs}--\ref{eq:exp_LEPA}), we find that visible effects in $\mu e$ scattering are excluded for $\Lambda \gtrsim 40~$GeV. Moreover, very stringent bounds on $g^e_Xg^\mu_X$ are set by the LHC experiments for $10 \lesssim M_U \lesssim 50~$GeV, via the measurement of the branching fraction for $Z$ decays to four leptons (electrons or muons)~\cite{Rainbolt:2018axw}, and by the BaBar experiment for $M_U \lesssim 10~$GeV~\cite{TheBABAR:2016rlg}. As a result, we find that observable effects in $\mu e$ scattering induced by NP lying below the electroweak scale are very unlikely.

%%%%%%%%%%%%%%%%%%%%%%%%%%%%%%%%%%%%%%%%%%%%%%%%%%%%%%%%
\subsection{Heavy NP and leptonic dipoles}

\begin{figure}
	\centering
	\includegraphics[width=0.21\textwidth]{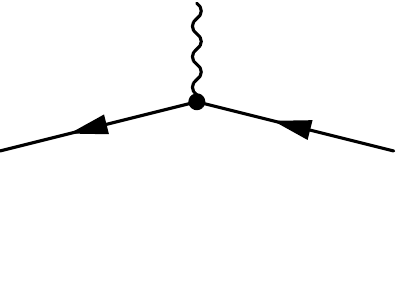}
	\hspace{0.7 truecm}
	\includegraphics[width=0.21\textwidth]{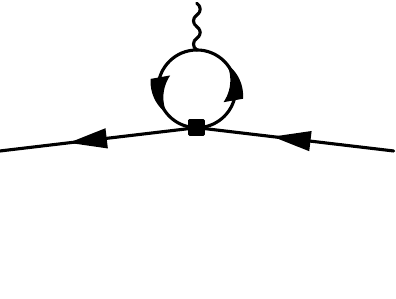}
	\vspace{-1.0 truecm}
\caption{Diagrams contributing to lepton dipole moments. On the left, the contribution from $\delta \mathcal{L}$,~\eq{L_eff_RGE}. On the right, the one-loop contribution from the four-lepton interactions contained in $\mathcal{L}_{\rm LFC}$,~\eq{LHEAVY_LFC}, denoted by a square.}
\label{fig1}
\end{figure}

The loop-induced Lagrangian $\delta \mathcal{L}_e$ of~\eq{L_eff_RGE} and the corresponding $\delta \mathcal{L}_{\mu}$ generate dipole moments for the electron and the muon.  Adding the contributions from $\delta \mathcal{L}_e$ to the one-loop diagrams shown in Fig.~(\ref{fig1}), we find the following contribution to the electron $g$-2
\begin{align}
	\Delta a_e &=  \frac{a_3}{2\pi^2} \, \frac{m M}{\Lambda^2} \log \left(\frac{\Lambda}{M}\right)
	\nonumber \\
	&\approx 10^{-12} \left( \frac{1\, {\rm TeV}}{\Lambda} \right)^{\!\!2} \!\! \left(\frac{a_3}{5 \times 10^{-2}}\right)\,,
\label{eq:Delta_a_e}
\end{align}
where $a_3 = a_T$, see Eq.~(\ref{eq:theas}). Comparing the SM prediction $a^{\rm{SM}}_e$~\cite{gm2eth} with the experimental measurement $a^{\rm{EXP}}_e$~\cite{Hanneke:2008tm} leads to $\Delta a_e = a^{\rm{EXP}}_e - a^{\rm{SM}}_e = (-88\pm 36) \times 10^{-14}$, where the latest atomic physics measurement of the fine-structure constant was employed~\cite{Parker:2018vye}. NP contributions are therefore allowed up to $|\Delta a_e|\lesssim 10^{-12}$~\cite{Giudice:2012ms,Crivellin:2018qmi}. The Lagrangian $\delta \mathcal{L}_e$ also generates an electron EDM
\begin{align}
	d_e &=  e \, \frac{a_4}{4\pi^2} \, \frac{M}{\Lambda^2} \log \left(\frac{\Lambda}{M}\right)
	\nonumber \\
	&\approx 10^{-29} \left( \frac{1\, {\rm TeV}}{\Lambda} \right)^{\!\!2} \!\! \left(\frac{a_4}{2 \times 10^{-8}}\right) ~ e \, {\rm cm},
\label{eq:d_e}
\end{align}
to be compared with the experimental bound $d^{{\rm{EXP}}}_e \!\leq\!~1.1 \!\times\! 10^{-29}~e$ cm~\cite{Andreev:2018ayy}. 
Notice that after adding the two contributions for the leptonic dipoles shown in Fig.~(\ref{fig1}), the dependence on the renormalization scale $\mu$ cancels, as physical observables are renormalization scale independent quantities. The predictions for $\Delta a_\mu$ and $d_\mu$ can be obtained from Eqs.~(\ref{eq:Delta_a_e}) and (\ref{eq:d_e}), respectively, via the replacement $M \leftrightarrow m$. As a result, after taking into account the current experimental bounds on $\Delta a_e$ and $d_e$, it turns out that $\Delta a_\mu$ and $d_\mu$ are irrelevant. The bound $|\Delta a_e|\lesssim 10^{-12}$ therefore prevents any possible NP contamination in $\mu e$ scattering arising from tensor interactions, as shown by Eqs.~(\ref{eq:bound_deltaLFC_aS_aT}) and (\ref{eq:Delta_a_e}).

%%%%%%%%%%%%%%%%%%%%%%%%%%%%%%%%%%%%%%%%%%%%%%%%%%%%%%%%
\subsection{Heavy NP and LFV effects}

We are now ready to analyse the phenomenological implications of the LFV Lagrangian of~\eq{LHEAVY_LFV}. As we already pointed out in Sec.~\ref{section:effectivelagrangian}, this Lagrangian violates the electron and muon family numbers by two units, thus generating, for example, the processes $e^-e^- \to \mu^-\mu^-$ and $\mu^+ e^- \to \mu^- e^+$. While no relevant experimental bounds are available for these two processes, tight constraints are set by the exclusion limits on muonium-antimuonium oscillation ${\rm Mu} \! - \! \overline{{\rm Mu}}$ (${\rm Mu}$ and $\overline{{\rm Mu}}$ are the $\mu^+ e^-$ and $\mu^- e^+$ bound states), a phenomenon predicted by Pontecorvo in 1957~\cite{Pontecorvo:1957cp} well before the discovery of the muonium in 1960~\cite{Hughes:1960zz}.

As shown by Feinberg and Weinberg~\cite{Feinberg:1961}, the time integrated probability for ${\rm Mu} \! - \! \overline{{\rm Mu}}$ oscillation reads
\be
	P\left( \overline{{\rm Mu}} \! - \! {\rm Mu}\right) \simeq 
	\frac{2\, |\langle \overline{{\rm Mu}}|\mathcal{L}_{\rm{LFV}}|{\rm Mu} \rangle|^2}{\Gamma_{\mu}^2},
\ee
where $\Gamma_{\mu} \!=\! m^5_\mu G^2_F/192\pi^3 \!\approx\! 3\times 10^{-10}~$eV is the muon decay rate. The evaluation of the (singlet) matrix element gives~\cite{Hou:1995dg} 
\be
	|\langle \overline{\rm Mu}|\mathcal{L}_{\rm{LFV}}|{\rm Mu} \rangle| = 
	\frac{| b_1 + b_2 -  3 b_3 - 4b_4 - 4b_5 |}{2\pi a_0^3 \Lambda^2},
\ee
where $a_0$ is the Bohr radius. The oscillation probability therefore reads
\be
	P\left(\overline{{\rm Mu}} \! - \! {\rm Mu} \right) \approx 
	10^{-9} \! \left( \!\frac{1 {\rm TeV}}{\Lambda} \!\right)^{\!\!4} \! |b_1 + b_2 -3b_3 - 4b_4 - 4b_5|^2,
\label{eq:probability_th}
\ee
to be compared with the current 90\%C.L.\ experimental bound~\citep{Willmann:1998gd} 
\be
	P\left(\overline{{\rm Mu}} \! - \! {\rm Mu} \right) \leq 8.2 \times 10^{-11}.
\label{eq:muonium_exp}	
\ee
Barring accidental cancellations in \eq{probability_th}, we obtain the following upper bounds
\begin{equation}
	|b_{l}| \lesssim 0.1 \left( \! \frac{\Lambda}{1 {\rm TeV}} \! \right)^{\!2}\,.
\label{eq:bound_muonium}
\end{equation}
In order to check whether the bounds of \eq{bound_muonium} allow visible effects in $\mu e$ collisions at MUonE, we computed the dominant contributions to the $\mu^+ e^- \to \mu^- e^+$ differential cross section. In the limit $|t|\gg M^2$ we find
\begin{align}
	\frac{d \sigma_{\rm {LFV}}}{dt}  =  & \,\, \frac{d \sigma_0}{dt} \, \delta_{\rm {LFV}}, 
	\\
	\delta_{\rm {LFV}} \approx & \left(\frac{t}{8 \pi \alpha \Lambda^2} \right)^{\!2}
    	\sum^{5}_{l=1} \left|b_l \right|^2  \! c_l(s,t),
\label{eq:delta_LFV}
\end{align}
where $b_l$ are the coefficients entering the Lagrangian $\mathcal{L}_{\rm LFV}$ of \eq{LHEAVY_LFV}, while the dimensionless functions $c_{l}(s,t)$ are, at MUonE, at most of order $\mathcal{O}(10)$. Since NP contaminations of MUonE's measurements can only occur for $\delta_{\rm {LFV}} \gtrsim 10^{-5}$, we end up with the following conditions,
\begin{align}
	\left| b_{l} \right|	\gtrsim  10^{3} \! \left( \! \frac{\Lambda}{1 {\rm TeV}} \! \right)^{\!2},
\label{eq:delta_LFV2}
\end{align}
which are excluded by the experimental bounds on muonium-antimuonium oscillation by orders of magnitude, see \eq{bound_muonium}.
As a result, we conclude that the extraction of $\Delta\alpha_{\rm h}(t)$ at the MUonE experiment will not be affected by LFV effects.

%%%%%%%%%%%%%%%%%%%%%%%%%%%%%%%%%%%%%%%%%%%%%%%%%%%%%%%%
\section{Light new physics mediators} \label{Light}
%%%%%%%%%%%%%%%%%%%%%%%%%%%%%%%%%%%%%%%%%%%%%%%%%%%%%%%%

In this Section we consider the impact of light mediators, with masses of order $\lesssim 1{\rm GeV}$, on the low-energy $\mu e$ scattering differential cross section. Since an EFT approach is not justified for this scenario, we proceed by specifying the spin and the interactions of the mediators with the SM particles. In particular, we discuss two benchmark scenarios with spin-0 or spin-1 dynamical particles. The former includes the case where the SM is supplemented by axionlike particles (ALPs), whereas the latter is representative of models with light dark photons or $Z^\prime$ vector bosons.

%%%%%%%%%%%%%%%%%%%%%%%%%%%%%%%%%%%%%%%%%%%%%%%%%%%%%%%%
\subsection{Light axionlike particles}

The most general Lagrangian describing the interactions of a spin zero particle $\Phi$ with leptons $\ell \!=\! (e,\mu)$ is
\begin{align}
	{\cal L}_\Phi &=
	\bar{\ell}_{L} C^{\Phi}_{\!R}  \ell_{R} \, \Phi + 
	\bar{\ell}_{R} C^{\Phi}_{\! L}  \ell_{L} \, \Phi + h.c.\,,
\label{eq:Lagrangian_ALP}
\end{align}
where $C^{\Phi}_{R}$ and $C^{\Phi}_{L}$ are flavor off-diagonal complex matrices which generally induce LFV~\cite{Bauer:2019gfk} 
and CP violating effects. If we further impose that ${\cal L}_\Phi$ is both flavor and CP conserving, we recover the standard expression
\begin{align}
	{\cal L}_{s+a} &= y_{s\ell} \,\bar{\ell} \ell \, s + i y_{a\ell}\, \bar{\ell} \gamma_5 \ell \, a\,,
\label{eq:Lagrangian_ALP_LFC}
\end{align}
where $s\,(a$) is a real scalar (pseudoscalar) field, while $y_{s\ell} = {\rm Re}(C^s_R+C^s_L)_{\ell \ell}$ and 
$y_{a\ell} = {\rm Im}(C^a_R-C^a_L)_{\ell \ell}$.

The $\mu e$ scattering process receives both LFC corrections from the $t$-channel exchange of an $s$ or $a$ particle, as well as LFV contributions from $t$- and $s$-channel exchanges. The former provide the following shifts of the $\mu e$ differential cross section
\begin{align}
	\delta_{\rm LFC}^{\pm, s} & =  \left( \! \frac{t}{8 \pi \alpha} \! \right)^{\!2}
     	\frac{(y_{s e} y_{s \mu})^2}{(t-m^2_s)^2} \frac{(t-4m^2)(t-4M^2)}{f(s,t)}   
    		\nonumber \\
	& \pm \frac{y_{s e} y_{s \mu}}{2 \pi \alpha} \frac{mM}{(t-m^2_s)} \frac{ t^2-2t ( m^2 + M^2 - s) }{ f(s,t) },       
\label{eq:delta_ALPs} \\
	\delta_{\rm LFC}^{\pm, a} & =  \left(\! \frac{t}{8 \pi \alpha} \! \right)^{\!2}
	\frac{(y_{a e} y_{a \mu})^2}{(t-m^2_a)^2} \frac{t^2}{f(s,t)}.  
\label{eq:delta_ALPa}
\end{align}
We note that the interference term with the leading QED contribution is present only when a scalar particle is exchanged, as already pointed out in Sec.~\ref{section:mue_scattering}. The most stringent bound on $y_{\Phi e}$, with $\Phi = s,a$, is set by the electron $g$-2 and reads 
\begin{align}
	|y_{\Phi e}| \lesssim 5 \times 10^{-4} \left( \frac{m_\Phi}{0.1\, {\rm GeV}} \right)\,,
\label{eq:bound_y_ePhi}
\end{align}
where we assumed that $m_e \ll m_\Phi$ and $|\Delta a_e|\lesssim 10^{-12}$. On the other hand, the muon $g$-2 as well as other low-energy experimental constraints impose the limit~\cite{Batell:2016ove}
\begin{align}
	|y_{\Phi \mu}| \lesssim 2 \times 10^{-3}\,,
\label{eq:bound_y_muPhi}
\end{align}
for $m_\Phi$ smaller than a few GeV. The combination of Eqs.~(\ref{eq:bound_y_ePhi}) and (\ref{eq:bound_y_muPhi}) therefore implies that 
	$y_{\Phi e} y_{\Phi \mu}\lesssim 10^{-6} (m_\Phi / 0.1\, {\rm GeV})$.
This bound leads to values of $\delta_{\rm LFC}^{\pm, \Phi}$ in Eqs.~(\ref{eq:delta_ALPs}) and (\ref{eq:delta_ALPa}) which are a few orders of magnitude below the resolution expected at MUonE. Moreover, we have checked that loop-induced vertex corrections arising from the couplings of~\eq{Lagrangian_ALP} are always irrelevant after imposing the above experimental  bounds. In principle we should also consider the couplings of $\Phi$ with photons. However, they contribute to $\mu e$ scattering only at loop level, and their effects are very suppressed once the electron and muon $g$-2 bounds are imposed~\cite{Marciano:2016yhf}.

Therefore, for light scalars and pseudoscalars with masses smaller than $\mathcal{O}(1{\rm GeV})$, the above LFC effects are not visible at the MUonE experiment.

LFV effects are constrained by the experimental limit on muonium-antimuonium oscillation, see~\eq{muonium_exp}, which implies the bound $|(C^{\Phi}_{X})_{\ell \ell^\prime}|\lesssim 10^{-4}(m_\Phi / 0.1\, {\rm GeV})$, where $X=L,R$. The resulting $\delta_{\rm LFV}^{\pm, \Phi}$ lies several orders of magnitude below the resolution expected at MUonE.

%%%%%%%%%%%%%%%%%%%%%%%%%%%%%%%%%%%%%%%%%%%%%%%%%%%%%%%%
\subsection{Light dark photons}

Models with extra $U(1)$ gauge groups are among the minimal and most studied extensions of the SM. In the so-called dark photon (DP) scenario, there is a kinetic mixing between the SM electromagnetic $U(1)_{\rm em}$ and the new $U(1)$. The relevant Lagrangian reads
\begin{align}
	{\mathcal  L}_{ \rm DP} = -\frac{1}{4}V_{\mu\nu}V^{\mu\nu}  + \frac{1}{2} m_{\rm {DP}}^2 V_\mu V^\mu 
	+ e \, \epsilon  J^{\mu}_{\rm em}V_\mu,
\label{LDP}
\end{align}
where $V^{\mu}$ and $V^{\mu\nu}$ are the DP field and field strength, $\epsilon$ is a dimensionless parameter accounting for the kinetic mixing, $m_{\rm {DP}}$ is the DP mass, and $J^{\mu}_{\rm em}$ is the electromagnetic current. The shift of the $\mu e$ differential cross section induced by $ {\mathcal  L}_{ \rm DP}$ is simply
\begin{align}
	\delta_{\rm {DP}}  \, = \, \frac{2\epsilon^2 t}{t-m^2_{\rm {DP}}}.
\label{eq:delta_DP}
\end{align}
For $m_{\rm {DP}} \lesssim 1$~GeV, the experimental limit on the kinetic mixing is $\epsilon^2  \lesssim 2 \times 10^{-7}$~\cite{Bauer:2018onh,Alemany:2019vsk}. Therefore, for dark photons with masses smaller than $\mathcal{O}(1{\rm GeV})$ the upper bound is $\delta_{\rm {DP}}  \lesssim 4 \times 10^{-7}$, a value well below the precision expected at MUonE. Moreover, this maximum $\delta_{\rm {DP}}$ value is obtained in the $m^2_{\rm {DP}} \ll |t|$ limit, when the $\delta_{\rm {DP}}$ correction becomes constant and therefore, as explained in Sec.~\ref{SM}, undetectable at the MUonE experiment. This point, namely the fact that MUonE will measure the shape of the differential $\mu e$ scattering cross section rather than its absolute value, was overlooked in Ref.~\cite{Schubert:2019nwm}. As a result, the authors of~\cite{Schubert:2019nwm} reached the incorrect conclusion that a very light dark photon can potentially affect MUonE's measurements.

%%%%%%%%%%%%%%%%%%%%%%%%%%%%%%%%%%%%%%%%%%%%%%%%%%%%%%%%
\subsection{Light \boldmath $Z^\prime$ vector bosons}

Models with an underlying $U(1)$ symmetry allow also direct couplings of SM particles with a new $Z^\prime$ vector boson. In particular, 
we discuss here the case of a massive $Z^\prime$ vector boson with mass $m_{Z^\prime}$ and leptonic couplings 
\begin{align}
	\mathcal{L}_{Z^\prime} =  g_V^\ell \, \bar\ell \gamma^\mu \ell \, Z^\prime_\mu + 
	g_A^\ell \, \bar\ell \gamma^\mu \gamma_5 \ell \, Z^\prime_\mu,
\end{align}
where $\ell = (e,\mu)$. The above Lagrangian induces the following leading shift of the $\mu e$ elastic cross section
\begin{align}
	\delta^{\pm}_{Z^\prime} & =  \frac{t}{2 \pi \alpha \, ( t - m^2_{Z^\prime} )} 
	 \bigg[ g^e_V g^\mu_V  + \, g^e_A g^\mu_A \, \frac{ t \left(z^{\pm} \!-\! z^{\mp} \right)}{2f(s,t)}
	\bigg],
\label{eq:deltaLFCZprime}
\end{align}
where $z^+ \!= \! s$ and $z^- \! =\! u$. We note that, contrary to the ALP scenario, the interference terms in~\eq{deltaLFCZprime} are not suppressed by the electron and muon masses and, therefore, the quadratic terms have been safely neglected. In contrast to the DP scenario, the constraints from hadronic colliders, beam dump and fixed target experiments are avoided by the Lagrangian $\mathcal{L}_{Z^\prime}$, and larger NP effects in $\mu e$ scattering can therefore be expected. The most stringent experimental bounds on the product of the vector couplings $g^e_V g^\mu_V$ are due to the light vector boson searches in the reaction $e^+e^- \to \gamma Z^\prime \to \ell^+\ell^- \gamma$ (with $\ell=e,\mu$) at BaBar~\cite{Lees:2014xha} and KLOE~\cite{Anastasi:2018azp}. Although these analyses were performed in the context of DP scenarios, to a good approximation they also apply to our framework~\cite{LemLm}. In particular, for $0.2 \lesssim m_{Z^\prime}\lesssim 10~$GeV, it turns out that $g^e_V g^\mu_V \lesssim 10^{-7}$, whereas in the range $0.02 \lesssim m_{Z^\prime}\lesssim 0.2~$GeV, below the dimuon mass treshold, only the $e^+e^-$ couplings can be constrained to $(g^e_V)^2 \lesssim 10^{-7}$~\cite{Lees:2014xha}. From this information we can conclude that the shift induced by a new $Z^\prime$ boson with purely vector couplings to electrons and muons, and mass in the range $0.2 \lesssim m_{Z^\prime}\lesssim 10~$GeV, is $\delta^{\pm}_{Z^\prime} \lesssim  \mathcal{O}(10^{-6})$, i.e.\ below MUonE's sensitivity. Further constraints on the couplings $g^\ell_{V,A}$ are set by the leptonic $g$-2. Indeed, imposing the conditions $|\Delta a_e| \lesssim 10^{-12}$, $|\Delta a_\mu| \lesssim 10^{-8}$ (in our numerical analysis we use the full one-loop expressions of Ref.~\cite{Leveille:1977rc}) and the BaBar limits~\cite{Lees:2014xha}, we find that $\delta^{\pm}_{Z^\prime} \lesssim 10^{-5}$ for $m_{Z^\prime}\lesssim 0.2$~GeV. 
Moreover, we have checked that loop-induced vertex corrections are always irrelevant after imposing the above experimental  bounds.

Possible effects induced by LFV $Z^\prime$ couplings to leptons are constrained by the experimental limits on muonium-antimuonium oscillation and the electron $g$-2. The resulting shift of the $\mu e$ differential cross section lies several orders of magnitude below MUonE's expected resolution.

Contrary to the light vector boson searches discussed above, to the best of our knowledge, searches of light $Z^\prime$ bosons with axial-vector couplings have not been performed so far. However, we expect that the bounds on $g^e_V g^\mu_V$ obtained in Refs.~\cite{Lees:2014xha,Anastasi:2018azp} apply, to a good approximation, also to $g^e_A g^\mu_A$.

%%%%%%%%%%%%%%%%%%%%%%%%%%%%%%%%%%%%%%%%%%%%%%%%%%%%%%%%
\section{Conclusions}
%%%%%%%%%%%%%%%%%%%%%%%%%%%%%%%%%%%%%%%%%%%%%%%%%%%%%%%%

The MUonE experiment has been proposed at CERN to determine the hadronic contribution to the effective electromagnetic coupling $\Delta \alpha_{\rm h}(q^2)$ from precise $\mu e$ elastic scattering data. In turn, this spacelike determination of $\Delta \alpha_{\rm h}(q^2)$ will yield a new precise prediction for the leading hadronic contribution to the muon $g$-2. This new, clean and inclusive approach, alternative to the traditional dispersive method based on low-energy hadronic $e^+e^-$ annihilation data, raises the question whether possible NP effects could contaminate MUonE's measurements at the level of its expected experimental resolution of $\mathcal{O} \left(10^{-5} \right)$.

We investigated this problem considering possible signals of NP in $\mu e$ collisions due to heavy or light mediators, depending on whether their mass is higher or lower than ${\cal O} (1{\rm GeV})$, the energy scale of the MUonE experiment. Heavy NP was analyzed by means of an effective field theory approach, considering the most general effective Lagrangian invariant under the electromagnetic gauge group and containing operators up to dimension-six. We computed the full set of NP corrections to the $\mu^{\pm} e^- \!\to\! \mu^{\pm} e^-$differential cross section, finding contributions from scalar, tensor, vector and axial-vector interactions. Scalar and tensor effects were found to be highly suppressed by the electron and muon masses, and could only contaminate MUonE's precise measurements for exceedingly large Wilson coefficients which are excluded by perturbativity and unitarity bounds. Moreover, tensor interactions would generate additional contributions to the electron and muon dipole moments, which are severely constrained experimentally. Possible contaminations at MUonE from scalar and tensor interactions are therefore excluded. Pseudoscalar interactions do not interfere with the leading QED amplitude and, therefore, do not contribute at the level of dimension-six operators. On the other hand, vector and axial-vector interactions could in principle generate detectable effects for smaller and, therefore, more plausible Wilson coefficients. However, via an explicit calculation of the NP corrections to the total cross section and forward-backward asymmetry of the process $e^+e^- \!\to \mu^+\mu^-$, we showed that existing experimental bounds disfavor any observable effect in $\mu^{\pm} e^- \!\to\! \mu^{\pm} e^-$ at MUonE. Our studies included possible LFV effects in $\mu e$ collisions generated by the process $\mu^+ e^- \to \mu^- e^+$. These effects were shown to be negligible at MUonE, as they would otherwise induce muonium-antimuonium oscillations beyond the present exclusion limits.

Light NP was analyzed specifying the spin and the interactions of the mediators with the SM particles. In particular, we studied benchmark scenarios with spin-0 and spin-1 dynamical particles. First, we focused on ALPs scenarios where the SM is supplemented by light (pseudo)scalars interacting with the leptons. For lepton flavor conserving interactions, we found that the bounds arising from the electron and muon $g$-2 force NP effects in $\mu^\pm e^- \!\to\! \mu^\pm e^-$ to lie well below MUonE's expected sensitivity. Similar conclusions can be reached for LFV interactions, as the existing experimental bounds on muonium-antimuonium oscillation provide once again formidable constraints. We then moved to the so-called dark photon (DP) scenario, where a new light vector boson interacting with the SM photon through a kinetic mixing is introduced. We showed that the present experimental limits on the DP kinetic mixing restrict NP effects in $\mu^{\pm} e^- \!\to\! \mu^{\pm} e^-$ well below MUonE's expected resolution. Finally, we considered the scenario with light $Z^\prime$ vector bosons interacting only with electrons and muons. Even in this less constrained case, possible contaminations of MUonE's measurements are disfavored by the present exclusion limits set by direct searches and leptonic dipole moments.

In conclusion, we showed that it is very unlikely that NP contributions will contaminate MUonE's extraction of $\Delta \alpha_{\rm h}(q^2)$ from the measurement of the $\mu e$ scattering differential cross section. The physics case of the MUonE proposal is therefore confirmed and reinforced by the present study.

{\em Note added.} After the completion of this work, we became aware of Ref.~\cite{Dev:2020drf} which addresses the sensitivity of MUonE to new light scalar or vector mediators able to explain the muon $g$-2 discrepancy. We reach similar conclusions on the points where our analyses overlap.

%%%%%%%%%%%%%%%%%%%%%%%%%%%%%%%%%%%%%%%%%%%
\subsection*{Acknowledgements}
%%%%%%%%%%%%%%%%%%%%%%%%%%%%%%%%%%%%%%%%%%%%%%%%%%%%%%%%
We thank G.~Abbiendi, L.~Calibbi, C.M.~Carloni Calame, E.J.~Chun, F.~Feruglio, G.F.~Giudice, J.~Herms, A.~Ibarra, J.~Jaeckel, P.~Mastrolia, F.~Piccinini, M.~Porrati, G.M.~Pruna, O.~Sumensari, R.~Torre, G.~Venanzoni and A.~Wulzer for fruitful discussions. We are also grateful to all our MUonE colleagues for our stimulating collaboration. 
We acknowledge partial support by FP10 ITN Elusives (H2020-MSCA-ITN-2015-674896) and Invisibles-Plus (H2020-MSCA-RISE-2015-690575). 
A.M.\ acknowledges the research grant ``The Dark Universe: A Synergic Multimessenger Approach", number 2017X7X85K under the program PRIN 2017 funded by MIUR, and thanks the Institut Pascal at the Universit\'e Paris-Saclay and the Kavli IPMU, University of Tokyo, for the hospitality and interesting discussions during the ``Astro-Particle Workshop'' and ``DM Workshop", respectively. 

%%%%%%%%%%%%%%%%%%%%%%%%%%%%%%%%%%%%%%%%%%

\end{document}